\documentclass{appolb}
\usepackage{epsfig}

\newcommand{\newton}[2]%
{\left(\raisebox{-1ex}{$\stackrel{\textstyle #1}{#2}$}\right)}

\begin{document}
\title{Collective Properties of the Exactly Solvable 
	Model of Ion-Channel Assemblies in a Biological Cell Membrane}
\author{Adam Rycerz
\address{Marian Smoluchowski Institute of Physics, 
	Jagiellonian University, \\
	ul.\ Reymonta 4, 30-059 Krak\'ow, 
	Poland}
}
\maketitle
\begin{abstract}
The behaviour of a system of ion channels formed across the cell membrane 
is presented. 
The infinite number of channels with an {\em infinite coupling} is introduced 
first as a reference point for the detailed derivation 
of the thermal-equilibrium probability distribution and
the classification of the phase transitions. 
Fluctuations in a {\em finite} system are discussed next. 
We propose a~new, {\em step-like} model of the ion channel switching, 
for which we provide the analytical results. 
The relation of this model to experiment is also provided.
Finally, the master equation for the finite channel-number
membrane is analysed numerically with the help of an 
{\em exact-diagonalization} technique.
In particular, the decay-time of a {\em metastable} solution is estimated. 
The results do not agree with those obtained perturbationally,
the difference is explained by the proposed {\em frozen-diffusion} approach.
\end{abstract}

\section{Introduction}

All living cells are effectively separated from the environment by the lipids 
that constitute the cell membrane. 
The lipid bilayer formation and its dynamics have been recently
objects of an intensive molecular dynamics studies \cite{fell}.
However, ions such 
as sodium $\mbox{Na}^+$ or potassium $\mbox{K}^+$ can
interact with proteins in the cell membrane
that transport the ions across the membrane. 
{\em Ion channels} are one of the three types of 
proteins involved in the ion transport;
the others are the {\em pumps} and the {\em carriers}. Approximately,
an equal portion of the ionic current passes through each of these routes 
\cite{frac}. 

Classic experiments using the patch clamp technique 
showed an interesting stochastic behaviour
of the ion channels that is controlled by the voltage across 
the membrane \cite{clamp}. For example, this behaviour is crucial
for the phenomena of neural excitability \cite{hohu}.
The {\em fractal model} 
of ion channel kinetics was formulated \cite{fracde}
and widely discussed in relation 
to the {\em Markov model} \cite{marfrac}.
Recently, Kramers' diffusion theory has been considered as an
alternative to those models \cite{kram}.
One should also mention a significant improvement in the data analysis
\cite{hearth} 
namely, a new method was developed for the identification 
of deterministic structure in the time series, 
such as obtained from the patch clamp experiments, 
and successfully applied to the hearth disease and the epilepsy research.

In all above experimental and theoretical analyses, the kinetics of 
the single ion-channel was investigated. In this paper we discuss 
collective effects in both infinite-size system (Sec.~\ref{dete}), 
and realistic system (Sec.~\ref{stat}-\ref{pertu}) of a few-hundred  
coupled ion channels along the cell membrane. 
We discuss the so-called {\em step-like} empirical model 
of the ion-channel switching affected by the voltage across the membrane 
that is based on that proposed by Hodgkin and Huxley \cite{hohu}, 
but is much simpler
and approximately rationalises the experimental data for the potassium 
and the sodium channels \cite{twohig}.

\section{Model and the physical system}
\label{model}

\begin{figure}[b]
\centering
\epsfxsize=0.8\textwidth
\epsfbox{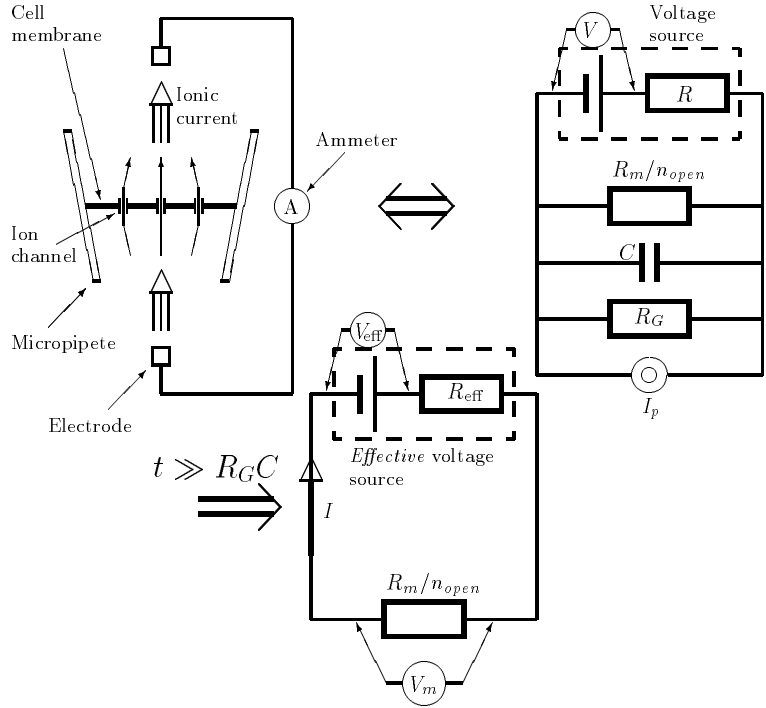}
\caption{The patch clamp experiment version for a
$m$-channel membrane and the corresponding electric circuit. 
The membrane parameters are: 
$I_p$ - the constant current of the ionic pumps,
$R_G$ - the resistivity for the ionic diffusion across the membrane,
$C$ - the membrane capacity,
and $R_m$ - the $m$-channel system resistivity when all
the channels open. For further considerations we suppose,
that the system is connected to the external voltage $V$ 
and resistivity $R$.
{\em Below:} the effective circuit for long-timescale processes
in the membrane ($t\gg R_GC$). 
The {\em effective} external voltage $V_{\rm eff}$
and resistivity $R_{\rm eff}$ are defined in the main text.}
\label{elec}
\end{figure}

In this section we discus a simple model of ion channel switching
and apply it to the system of $m$-channels across the membrane 
(see Fig.~\ref{elec}). 

The ion channels are formed by specific membrane proteins which 
undergo a spontaneous, voltage-sensing conformational transitions
between the {\em open} and the {\em closed} states \cite{hill}.
Hereinafter, we will concentrate on isolated assembly of $m$ ion
channels of one type, i.e.\ potassium K$^+$. 
Let us suppose the simplest 
$2$-state Markov model of the single channel switching
(this is not a strict supposition, because 
such a model is equivalent to the $m$+1~-state 
Markov model of the whole membrane, cf.\ Sec.~\ref{stat}). 
In this model such a channel is characterised
by two microscopic parameters determined 
by the voltage $V_m$ across the membrane:
the inverse average open- and closed times 
($\lambda_{open}=\lambda_{open}\left(V_m\right)$, 
and 
$\lambda_{closed}=\lambda_{closed}\left(V_m\right)$, respectively).
Physically, parameters $\lambda_{open}$ and $\lambda_{closed}$
are measured indirectly in a time series obtained from patch clamp
experiment for single channel \cite{clamp}, whereas
$V_m$ is the driving parameter controlled in such an experiment. 
Subsequently, the expectation value of the fraction 
of opened channels $\left<n_{open}\right>$
obey the two-state, ``open-closing'' dynamics \cite{hohu}:
\begin{equation}
  \label{nopenll}
  \frac{d}{dt}\left<n_{open}\right> =
  \lambda_{open}
  \left(1-\left<n_{open}\right>\right)-
  \lambda_{closed}
  \left<n_{open}\right>. 
\end{equation}
For the sake of convenience, one can define the decay constant 
$\lambda = \lambda_{open}+\lambda_{closed}$
(that provides natural timescale of the processes),
and the probability of single channel open 
$p_{open}=\lambda_{open}/\lambda$, 
which allow us to rewrite eq.~(\ref{nopenll}) in form:
\begin{equation}
  \label{nopenlp}
  \frac{1}{\lambda}\frac{d}{dt}
  \left<n_{open}\right> =
  p_{open}-\left<n_{open}\right>. 
\end{equation}
In further considerations, parameter $\lambda = \lambda\left(V_m\right)$
is regarded as constant, and used to define the dimensionless time:
\begin{equation}
  \label{ltt}
  \tau = \lambda t. 
\end{equation}
One can verify this approximation for a broad 
range of voltage across the membrane $V_m$ by looking at
the experimental data (see Table \ref{kexp}). 
With this simplification, only one microscopic
parameter $p_{open}\left(V_m\right)$ remains. 

\begin{table}[t]
\caption{Microscopic parameters of single 
potassium channel from human ocular epithelial cells 
(after Twohig, Ref. \cite{twohig}).}
\centering
\begin{tabular}{|c|c|c|c|c|}
\hline\hline
Membrane & Ave.\ open & Ave.\ closed & Time & Prob. of \\ 
potential & time & time & unit & open chan. \\ 
$V_m$ [mV] & $1/\lambda_{open}$ [ms]
 &  $1/\lambda_{closed}$ [ms] & $1/\lambda$ [ms]
 & $p_{open}$ \\ \hline
 -20 & 14.57 & 1.58 & 1.425 & 0.098 \\ 
 -30 & 11.84 & 2.04 & 1.740 & 0.147 \\ 
 -40 & 5.18 & 3.62 & 2.131 & 0.411 \\ 
 -50 & 5.42 & 5.42 & 2.710 & 0.500 \\ 
 -60 & 3.94 & 8.06 & 2.646 & 0.672 \\ 
 -70 & 2.72 & 15.46 & 2.313 & 0.850 \\ 
 -80 & 1.77 & 39.37 & 1.694 & 0.957 \\ 
 -90 & 1.71 & 60.32 & 1.663 & 0.973 \\ \hline\hline
\end{tabular}
\label{kexp}
\end{table}

Our starting assumptions, concerning the electrical properties of
the membrane (see Fig.~\ref{elec}), are as follows:
$(i)$ the conductance of the membrane is directly proportional to the
number of open channels, i.e.\ its resistivity is equal to $R_m/n_{open}$, 
where $R_m$ is the resistivity when all the channels are open,
$(ii)$ the total current of the ionic pumps $I_p$ is constant, 
$(iii)$ the voltage-induced ionic diffusion across the membrane is described
by the linear resistivity $R_G$, 
$(iv)$ the relevant time-scale is that of  long-time processes $t\gg R_GC$, 
where $C$ is the membrane capacity, 
and $(v)$ whole the system $(i)$-$(iv)$ is coupled to the constant voltage $V$ 
source with internal resistivity $R$.

After introducing all above suppositions, a direct application
of the Kirchoff's laws to the electric circuit shown in Fig.~\ref{elec} 
proves its equivalence  to the {\em effective} circuit, which contains only 
an assembly of switching channels coupled to the external
voltage $V_{\rm eff}$ and the resistivity $R_{\rm eff}$:
\begin{equation}
  \label{veff}
  V_{\rm eff}=\frac{V-RI_p}{1-R/R_G},\ \ \ \ \ \ 
  R_{\rm eff}=\frac{R}{1-R/R_G}.
\end{equation}
The calculation of the relation between the
voltage across the membrane $V_m$ and the fraction of 
open  channels $n_{open}$ for the {\em effective} electric circuit
is straightforward and yields
\begin{equation}
  \label{vmv}
  V_m = \frac{V_{\rm eff}}{1+n_{open}R_{\rm eff}/R_m}.
\end{equation}
This allows us to consider the probability of a single channel open 
as a function $p_{open}=p_{open}\left(n_{open}\right)$,
since $V_m$ is a function of $n_{open}$ (\ref{vmv}). 
Moreover, instead of applying phenomenological models of single channel
switching $p_{open}\left(V_m\right)$, such that proposed by Hodgkin 
and Huxley \cite{hohu}, we can now consider indirectly a model function 
$p_{open}\left(n_{open}\right)$, describing the assembly of coupled ion
channels themselves. 

\begin{figure}[b]
\unitlength=0.01\textwidth
\begin{picture}(100,40)
\put(0,-12){\epsfxsize=\textwidth\epsfbox{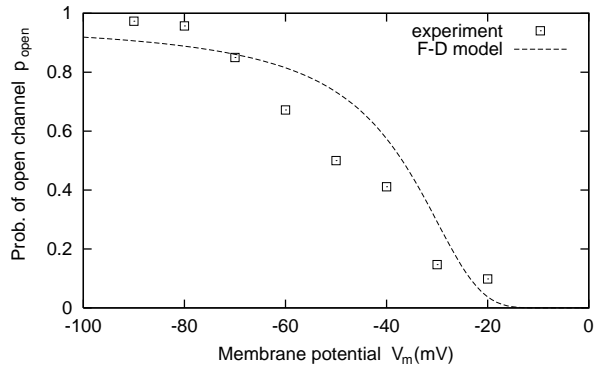}}
\end{picture}
\caption{Comparison between the {\em step-like}
model of ion channel switching for the fitted empirical
parameters $A=-3.85\pm 0.78$ and $V_0 = -142\pm 30\mbox{ mV}$ 
(see main text for a detailed description) 
and the experimental data for human ocular epithelial cells
listed in Table \ref{kexp}.}
\label{pvmexp}
\end{figure}

Namely, for a qualitative analysis of the membrane, 
one can select any model function $p_{open}\left(n_{open}\right)$
that retains principal properties of physical
system, such as {\em negative feedback} (which means
that channel remains closed for high polarising 
voltage $V_m$ and open for low, see \cite{twohig}),
and have convenient analytical properties.
In Sec.~\ref{stat}-\ref{pertu} we apply the {\em step-like} function
\begin{equation}
  \label{fedi}
  p_{open}\left(n_{open}\right) = 
  \left(1+e^{B\left(n_0-n_{open}\right)}\right)^{-1}, 
\end{equation}
where $B$ and $n_0$ are adjustable parameters, 
which allow us to perform some integrations analytically. 
A brief comparison of the model of channel switching
defined by the function (\ref{fedi}) with some of available experimental
data \cite{twohig} are provided in Fig.~\ref{pvmexp}.
Strictly speaking, the relation inverse to (\ref{vmv})
permitted us to write the probability of channel open as a function of
voltage again, i.e.\
\begin{equation}
  \label{fedivm}
  p_{open}\left(V_m\right) = 
  \left(1+e^{A+V_0/V_m}\right)^{-1},
\end{equation}
where $A=B(n_0+R_m/R_{\rm eff})$ and $V_0=-BV_{\rm eff}R_m/R_{\rm eff}$ 
are empirical parameters, constant for a particular type of 
ion channel (i.e. potassium channel for human ocular 
epithelial cells, see Fig. \ref{pvmexp}). 
Empirical parameters $A$ and $V_0$ provides the
relation between model parameters $B$, $n_0$ 
and the physical driving parameters $V_{\rm eff}$, $R_m/R_{\rm eff}$ 
(see Fig. \ref{elec}):
\begin{equation}
  \label{modphy}
  V_{\rm eff} = \frac{V_0}{Bn_0-A}, \ \ \ \ \ \ 
  R_m/R_{\rm eff} = A/B - n_0.   
\end{equation}

In effect, the selection of the probability of channel open in the form
(\ref{fedi}) constitutes the {\em step-like} model, which, as we just shown,
rationalises the experimental data of Ref.\ \cite{twohig}.

\section{Deterministic system: long-range coupled infinite membrane}
\label{dete}

In this section we consider the limiting case of the membrane 
with infinite number of channels ($m\rightarrow\infty$), and
far from a critical point \cite{landau}.  For such a system 
one can suppose that the probability distribution is narrow 
and, subsequently, the fraction of open channels 
$n_{open}$ is close to its average value 
$n_{open}\approx\left<n_{open}\right>$. With this assumption (which will be 
verified in the next section), we can write down an approximate  
version of Eq.~(\ref{nopenlp}) as follows
\begin{equation}
  \label{nopenpn}
  \frac{d}{d\tau}n_{open} = 
  p_{open}\left(n_{open}\right) - n_{open}, 
\end{equation}
where $\tau$ is a dimensionless time (\ref{ltt}). 
Equation (\ref{nopenpn}) becomes exact in limit $m\rightarrow\infty$
(except in the case when the system undergoes the 2-nd order phase
transition), as a {\em mean-field} solution of the systems with the infinite
coupling \cite{landau}.

\begin{figure}[b]
\unitlength=0.01\textwidth
\begin{picture}(100,40)
\put(0,-12){\epsfxsize=\textwidth\epsfbox{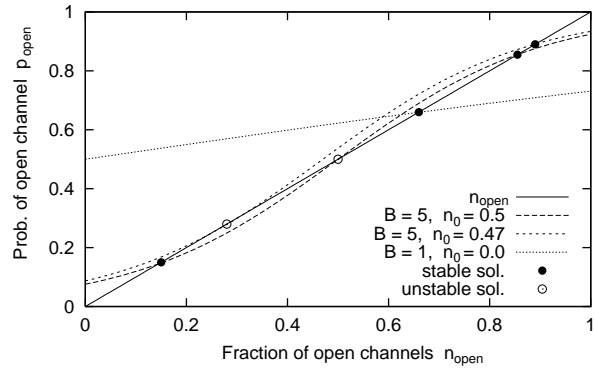}}
\end{picture}
\caption{Stable and unstable stationary solutions 
$n_* = p_{open}\left(n_*\right)$ 
of the dynamical equation for the infinite membrane (see main text).
The step-like model of ion channel switching is applied.}
\label{pnfedi}
\end{figure}

For all the models of ion channel switching \cite{twohig}, 
qualitatively similar to that defined by (\ref{fedi}), 
eq.~(\ref{nopenpn}) could have from one to three 
stationary solutions $n_{open}(\tau)=n_*$, that obey the relation
(cf.\ Fig.~\ref{pnfedi})
\begin{equation}
  \label{nopenstat}
  p_{open}\left(n_*\right) = n_*.
\end{equation}
For each of these solutions, one can consider its stability against small
fluctuations $n_{open}=n_*+\delta n$, by making 
a proper linearization of the relation (\ref{nopenstat}):
\begin{equation}
  \label{popenstat}
  p_{open}\left(n_{open}\right) = 
  p_{open}\left(n_*+\delta n\right) =
  n_* + p_*'\delta n + {\cal O}\left(\delta n^2\right),
\end{equation}
where $p_*'$ is the first derivative of the probability 
as a function of the fraction of open channels: 
$p_*'=p_{open}'\left(n_*\right)$. 
The equation (\ref{popenstat}), after substituting to the dynamical 
eq.~(\ref{nopenpn}) provides us with an approximate result 
(for $\delta n << 1$): 
\begin{equation}
  \delta n(\tau) = \delta n(0) e^{-\left(1-p_*'\right)\tau}. 
\end{equation}
This result simply means, 
that the first derivative of the probability $p_*'$
in the stationary point $n_*$ defined by the relation 
(\ref{nopenstat}), allows us to identify {\em stable} 
(for $p_*'< 1$) and {\em unstable} (for $p_*'< 1$) stationary
solutions of the dynamical eq.~(\ref{nopenpn}) 
(see Fig.~\ref{pnfedi}). The threshold value $p_*'=1$ will
be shown to correspond with the critical point in terms of 2-nd
order phase transition in the next section.

The approach presented above provides us two equivalent stable solutions 
(see Fig.~\ref{pnfedi}). 
In the next section we will start the considerations
for finite number of channels $m$, and will verify
(in Sec.~\ref{dyna}) one of these stable points to be actually
{\em metastable}, with a finite decay time growing exponentially
with $m$.

\section{Stationary probability distribution for a finite membrane}
\label{stat}

\begin{figure}[b]
\centering
\epsfxsize=0.8\textwidth
\epsfbox{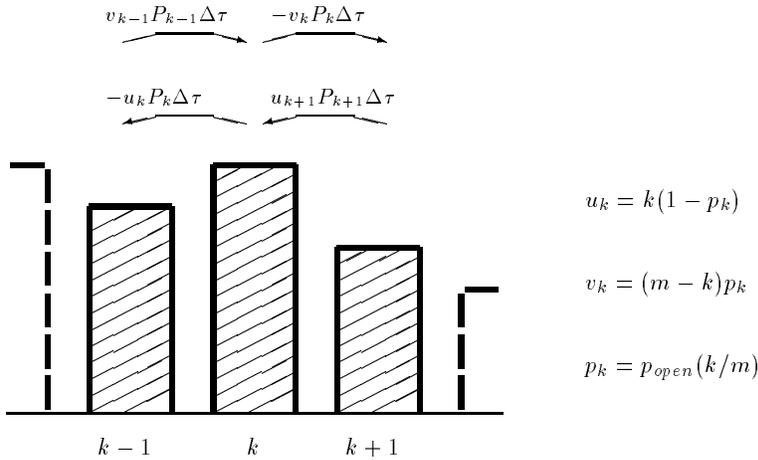}
\caption{Time evolution of the probability distribution for finite 
$m$-channel membrane toward the stationary state $\left\{P_k^{(0)}\right\}$. 
Index $k=0,\ldots,m$ is the number of open channels, 
the amounts of probability flowing between each pair
of the points in the dimensionless time interval $\Delta\tau$
are shown. 
The coefficients $u_k$, $v_k$ and $p_k$ in probability flows 
are defined above. }
\label{nflow}
\end{figure}

Let us consider the finite, $m$ ion channel cell membrane, 
and the evolution of its probability distribution 
$P=\left\{P_k\right\}$,
(where $k=0,\ldots,m$ is the number of open channels, and the 
normalization condition $\sum_kP_k=1$ is satisfied) in time. 
All the probability flows between $k$-th point 
and two neighbouring points are shown symbolically 
on Fig.~\ref{nflow}.

In the limit of the stationary probability distribution 
$P^{(0)}=\left\{P_k^{(0)}\right\}$
one can write down the detailed-balance equation between 
$k-1$ and $k$ points (cf.\ Fig.~\ref{nflow}) in the form
\begin{equation}
  \label{detbalnn}
  (m-k+1)p_{k-1}P_{k-1}^{(0)}-k(1-p_k)P_k^{(0)} = 0
\end{equation}
for each $k=1,\ldots,m$, 
where $p_k$ is the probability of open channel
when $k$ channels are open, 
$p_k = p_{open}(k/m)$ (see Sec.~\ref{model}). 
For non-zero probability distribution (this seems 
to be natural supposition for physical 
systems and will be verified {\em a posteriori})
one can rewrite Eq.~(\ref{detbalnn}) in the form
\begin{equation}
  \label{detbalr}
  P_k^{(0)} = \frac{m-k+1}{k}\frac{p_{k-1}}{1-p_k}P_{k-1}^{(0)},
\end{equation}
for $k=1,\ldots,m$, that determines the stationary probability
distribution $P^{(0)}=\left\{P_k^{(0)}\right\}$ except of the multiplicative
constant $P_0^{(0)}$, which has to be adjusted  to conserve
normalization $\sum_kP_k^{(0)}=1$. 
Eq.~(\ref{detbalr}) provides the powerful
tool for estimating numerically the stationary probability distribution for 
$m$-channel system with given model of ion channel switching $p_{open}(n)$.  
Moreover, in Appendix \ref{detbal}, using Eq.~(\ref{detbalr}),
we discuss small fluctuations around the stationary solutions (\ref{nopenstat})
for a general model $p_{open}(n)$.

\begin{figure}[b]
\unitlength=0.01\textwidth
\begin{picture}(100,70)
\put(0,19.1){\epsfxsize=\textwidth\epsfbox{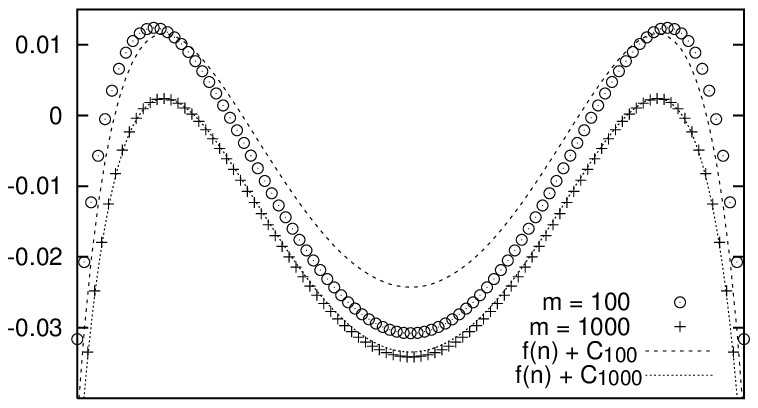}}
\put(0,-12){\epsfxsize=\textwidth\epsfbox{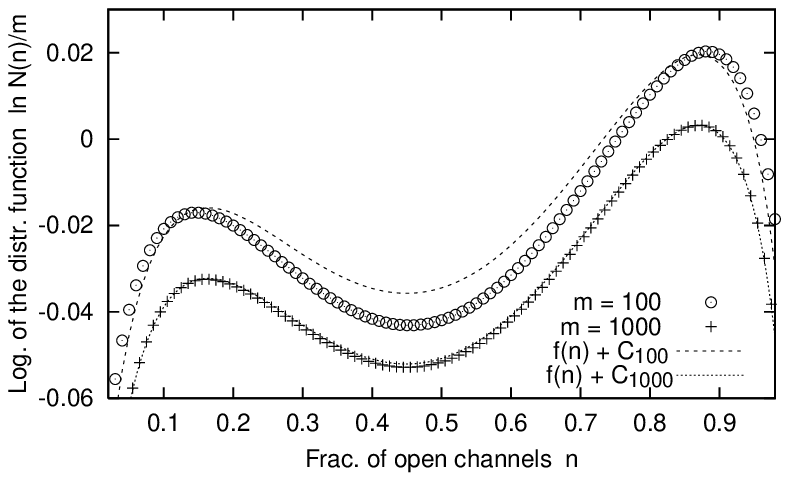}}
\end{picture}
\caption{Exact numerical solutions of the detailed-balance 
equations for different $m$ compared with 
an analytical solution of corresponding differential equation for
$m\rightarrow\infty$ (see main text).
The quasi-continuous probability distribution function 
is defined as $P(n)=mP_k$, where $n=k/m$. 
{\em Above}: symmetric step-like model of ion channel
switching ($B=5$, $n_0=0.5$), {\em below}: an asymmetric case
($B=5$, $n_0=0.49$)}
\label{stadis}
\end{figure}

The numerical results, presented in Fig.~\ref{stadis}, 
as well as an approximate formula for small 
fluctuations (\ref{ngauss}), derived in App.~\ref{detbal}, 
suggest the approximate analytical way
of solving detail balance equations for
large $m$. In this limit, one can 
look for universal function $f(k/m)$, such that
\begin{equation}
  \label{nexpf}
  P_0(n) = e^{m\left(f(n)+C_m\right)}. 
\end{equation}
In Eq.\ (\ref{nexpf}) we define a {\em quasi-continuous} variables:  
fraction of open channels $n=k/m$,  
and the stationary probability distribution function
$P_0(n)$, defined by relation $P_k^{(0)}=P_0(k/m)\Delta n$,
where $\Delta n = 1/m$. 
$C_m$ is an adjustable parameter, so that function 
$P_0(n)$ (\ref{nexpf})
satisfy the normalization condition $\int^1_0dnP_0(n)=1$. 
One can roughly estimate the value 
of $C_m$ by making Gaussian approximation 
around the global maximum of $f$: 
$f(n)\approx f_{max} - f_{max}''(n-n_{max})^2/2$, 
obtaining: 
\begin{equation}
  \label{cmapp}
  C_m= -f_{max} + \frac{1}{2}\frac{\ln m}{m} + {\cal O}(1/m).
\end{equation}
Then, making an expansion
\begin{equation}
  \label{mfexp}
  mf(n-\Delta n)=
  m\left[f(n)-\frac{df}{dn}\Delta n + 
  {\cal O}\left(\Delta n^2\right)\right]=
  mf(n) - \frac{df}{dn} + {\cal O}(1/m),
\end{equation}
and substituting it into the detail-balance equations (\ref{detbalr}),
we obtain:
\begin{equation}
  \label{edfdn}
  e^{-\frac{df}{dn}}=\frac{n}{1-n}
  \frac{1-p_{open}(n)}{p_{open}(n)}+{\cal O}(1/m). 
\end{equation}

\begin{figure}[b]
\unitlength=0.01\textwidth
\begin{picture}(100,70)
\put(0,19.1){\epsfxsize=\textwidth\epsfbox{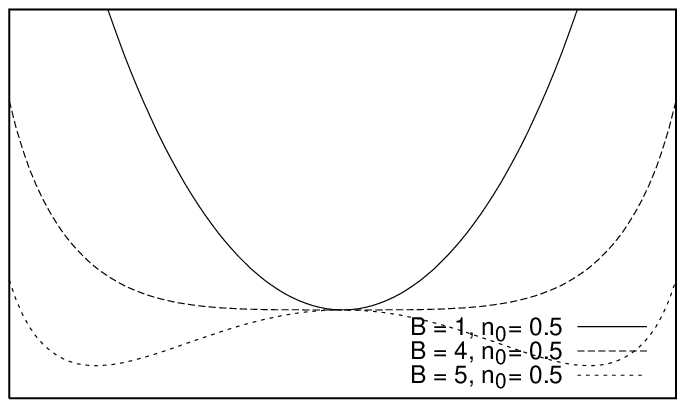}}
\put(0,-12){\epsfxsize=\textwidth\epsfbox{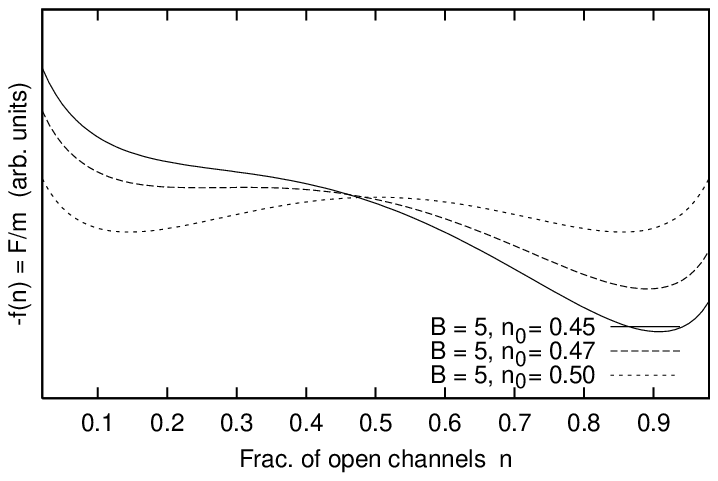}}
\end{picture}
\caption{Free energy per channel in the step-like
model of ion channel switching for different parameters
$B$ and $n_0$. {\em Top panel}: 2-nd order phase transition for
symmetric case ($n_0=0.5$) and varying $B$, {\em Bottom}:
the phase transition for fixed $B=5$ and varying $n_0$
({\em note} the presence of the secondary minimum on the left, 
corresponding to a metastable state).}
\label{phatra}
\end{figure}

Eq.\ (\ref{edfdn}) in turn, provides the approximate differential
equation for function $f(n)$ (exact in the limit $m\rightarrow\infty$):
\begin{equation}
  \label{dfdn}
  \frac{df}{dn} = \ln\left(\frac{1-n}{n}
  \frac{p_{open}(n)}{1-p_{open}(n)}\right).  
\end{equation}
In particular, when looking for extrema of the function $f(n)$ 
(the condition $df/dn=0$),
which are equivalent to the extrema of the stationary probability
distribution function $P_0(n)$ (\ref{nexpf}), 
Eq.~(\ref{dfdn}) reduces to the condition for stationary solutions 
of the deterministic equation of motion (\ref{nopenpn}) 
$n_*=p_{open}(n_*)$ (see Sec.~\ref{dete}).
This provides the link between these two approaches.  

Eq.~(\ref{dfdn}) can be solved analytically
for the step-like model of ion channel switching $p_{open}(n)$ 
defined by Fermi-Dirac function (\ref{fedi}):
\begin{equation}
  \label{fnfedi}
  f(n)= -n\ln n -(1-n)\ln (1-n)
  + \frac{B}{2}(n-n_0)^2.
\end{equation}
Plots of the function (\ref{fnfedi})
for sample parameters $B$ and $n_0$ are shown
in Fig.~\ref{phatra}. According to the formal 
correspondence of Eq.~(\ref{nexpf}) with 
the probability distribution for Gibbs canonical ensemble \cite{landau}
\begin{equation}
  \label{freeendef}
  P_k=\frac{Z_k}{\sum_kZ_k}=e^{-{\cal F}(n,B,n_0)/\Theta},
\end{equation}
where $Z_k$ is the sum of states above all $\newton{m}{k}$
configurations for given number of open channels $k$, 
$\Theta$ is the reduced temperature, and ${\cal F}(n,B,n_0)$
is the free energy of the system with given
fraction of open channels $n$, one can formally define 
the mean-field {\em Ginzburg-Landau} \cite{landau}
free energy functional for $m\rightarrow\infty$ limit
(and subsequently $n\approx\left<n\right>$):
\begin{equation}
  \label{freeen}
  {\cal F}(n,B,n_0)=-mf(n).
\end{equation}
Here we will concentrate only on the {\em isothermal} 
ensemble ($\Theta=1$), \cite{further}. 
The definition (\ref{freeendef}) allows us to identify the candidates 
for 1-st and 2-nd order phase transitions (cf.\ Fig.\ref{phatra}) 
in terms of Ginzburg-Landau theory \cite{landau}. 
However, when the parameter $B$ is constant and $n_0$ varies,
the configurational entropy of the system $S_{\rm conf}=\ln\newton{m}{k}$ 
does not change when the fraction of open channels $n$ switch between
two equivalent minima at the point $n=0.5$, 
so this process also has to be considered as the 2-nd order phase 
transitions within the Ehrenfest classification scheme.
The remarkable feature of this case is possible existence of a
metastable state (cf.\ Fig.\ref{phatra}), that motivated us 
to study the dynamics of the system in the next section.

\section{Dynamics of the probability distribution 
	via an exact diagonalization}
\label{dyna}

In this section we derive the system of dynamical equations 
(the {\em master equation}),
which describe time evolution of the probability distribution
of the number of open channels for the $m$-channel membrane. 
We will also present results obtained when solving this system 
by using the numerical {\em exact diagonalization} technique. 
These results are substantial for further considerations, 
i.e. they are used to justify the approximate approach proposed 
in Sec.~\ref{pertu} for large systems. 

Let us consider a discrete probability distribution for 
the $m$-channel membrane $\left(P_k\right)$, where $k=0,\ldots,m$ 
is the number of open channels, and the normalization condition
$\sum_kP_k=1$ is satisfied (see Sec.~\ref{stat}). 
Taking into account probability flows between the $k$-th point
and its two neighbours $k-1$, $k+1$ (see Fig.~\ref{nflow})
one can write down 
the system of linear, ordinary differential equations:
\begin{equation}
  \label{dnkt}
  \frac{d}{d\tau}P_k=-\sum_{l=0}^{m}A_{kl}P_l,
\end{equation}
where matrix $[A_{kl}]$ has a tridiagonal form
$
  A_{kl}=(u_k+v_k)\delta_{kl}
  - u_{k+1}\delta_{k+1,l} - v_{k-1}\delta_{k-1,l}, 
$
$\delta_{kl}$ is the Kronecker delta, $u_k=k(1-p_k)$, $v_k=(m-k)p_k$
(see also Fig.~\ref{nflow}), and the probabilities $p_k=p_{open}(k/m)$ 
are given by the model of ion channel switching (cf.\ Sec.~\ref{model}).

The system (\ref{dnkt}) can be simplified by the rescaling $L_k=P_k/\alpha_k$, 
where coefficients $\left(\alpha_k\right)$
are defined for $k=1,\ldots,m$ by the recursive formula:
\begin{equation}
  \label{alphak}
  \alpha_k=\sqrt{\frac{v_{k-1}}{u_k}}=
  \left(\frac{m-k+1}{k}\frac{p_{k-1}}{p_k}\right)^{1/2},
\end{equation}
and $\alpha_0$ remains a free multiplicative parameter. 
However, comparing the formula (\ref{alphak}) and the detailed balance 
Eq.~(\ref{detbalr}), one can choose  $\alpha_k=\sqrt{P_k^{(0)}}$ for
$k=0,\ldots,m$ (cf.\ Sec.~\ref{stat}), and adjust $\alpha_0$ appropriately. 
That form will be suitable for e.g.\ perturbation approach 
(see Sec.\ \ref{pertu}). 
In terms of variables $\left(L_k\right)$, the linear system (\ref{dnkt})
takes the form:
\begin{equation}
  \label{dlkt}
  \frac{d}{d\tau}L_k=-\sum_{l=0}^{m}B_{kl}L_l,
\end{equation}
where $[B_{kl}]$ is the symmetric, tridiagonal matrix
with diagonal elements $B_{kk} = u_k + v_k$, and off-diagonal
$ B_{k-1,k} = -\alpha_ku_k/\alpha_{k-1}  = -\sqrt{v_{k-1}u_k}.$
For particular model of ion channel switching, in our case the step-like
model (\ref{fedi}), 
the matrix $[B_{kl}]$ can be diagonalized with standard numerical packages
\cite{lapack}, up to $m\approx 1000$. 
Presented technique allows us to examine time intervals 
as long as $\tau\approx 10^{10}$ (\ref{ltt}), 
unreachable by an indirect numerical methods of 
solving of the system (\ref{dnkt}) (e.g. with Runge-Kutta 
method). These features have substantial meaning for searching
{\em metastable} states for a large system
that will be discussed further in this section.

The formal solution of the system (\ref{dnkt})
is given by the linear combination:
\begin{equation}
  \label{nsol}
  P_k(\tau) = \alpha_k\sum_{l=0}^m c_l
  L_k^{(l)} e^{-\omega_l\tau},
\end{equation}
where $\omega_l$ are the eigenvalues of $[B_{kl}]$
in ascending order $\omega_0\leq \ldots\leq\omega_m$, 
$\{L^{(l)}\}$ are the corresponding, orthonormal eigenvectors, 
$L^{(l)}=\left\{L_k^{(l)}\right\}$, $k,l=0,\ldots,m$
(so that $\sum_kL_k^{(l)}L_k^{(n)}=\delta_{ln}$),
and coefficients $c_l$ are defined by decomposition
of the initial probability distribution $\left\{P_k(\tau=0)\right\}$ 
in the basis $\{L^{(l)}\}$:
\begin{equation}
  \label{nsolck}
  c_l = \sum_{k=0}^mL_k^{(l)}P_k(0)/\alpha_k.
\end{equation}
One can find the lowest eigenvalue of the matrix $[B_{kl}]$
to be equal $\omega_0=0$, 
by substituting $L_k^{(0)}=\sqrt{P_k^{(0)}}$, for $k=0,\ldots,m$,
where $P^{(0)}=\left\{P_k^{(0)}\right\}$ is the stationary probability  
distribution obtained in Sec.~\ref{stat} from detailed-balance 
equation (\ref{detbalnn}). 
In other words, the solution (\ref{nsol}) of the system (\ref{dnkt})
relaxes to the stationary distribution $\left\{P_k^{(0)}\right\}$, 
for $\tau\rightarrow\infty$. 
In all the numerical analysis below, we examine the
details of this relaxation, particularly 
we search the possible {\em metastable}
state associated with the secondary maximum of the stationary
probability distribution $\left\{P_k^{(0)}\right\}$ (see Fig.~\ref{stadis}).

\begin{figure}[b]
\unitlength=0.01\textwidth
\begin{picture}(100,70)
\put(0,19.1){\epsfxsize=\textwidth\epsfbox{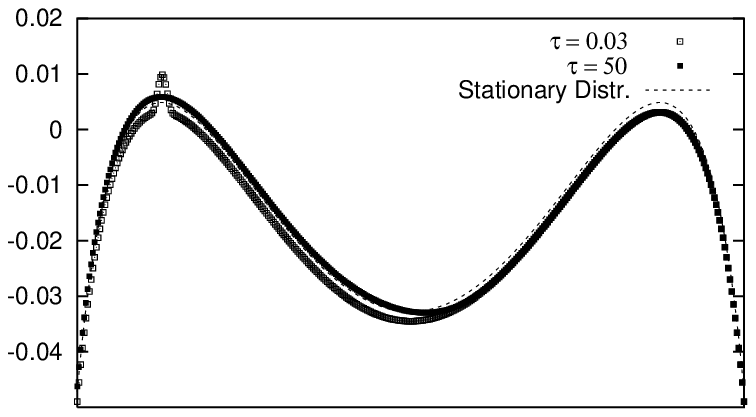}}
\put(0,-12){\epsfxsize=\textwidth\epsfbox{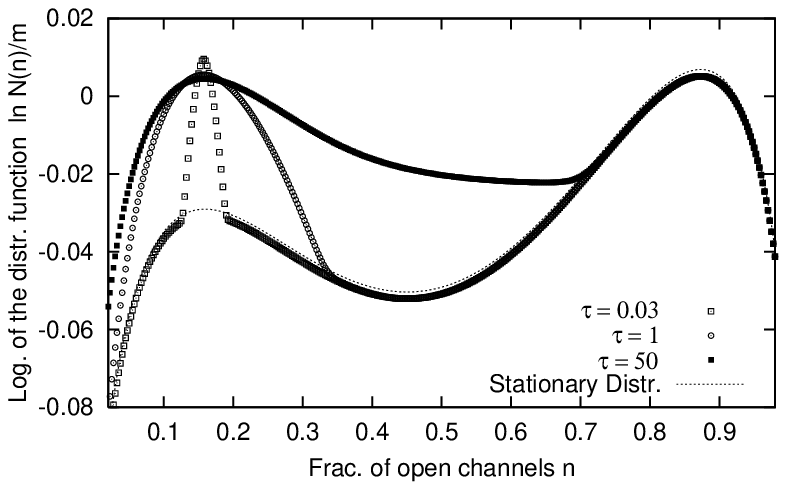}}
\end{picture}
\caption{
Evolution of the
initial probability distribution
for $m=400$ channels membrane (see main text for details)
to the metastable solution,  
obtained from the numerical {\em exact diagonalization} technique.
The stationary solution for $\tau\rightarrow\infty$ 
is also presented.
The quasi-continuous probability distribution function 
is defined as $P(n)=mP_k$, where $n=k/m$. 
{\em Top panel}: symmetric step-like model of ion channel
switching with $n_0=0.5$ and $B=5$.
{\em Bottom}: an asymmetric case
with $n_0=0.49$ and $B=5$.}
\label{stepi}
\end{figure}

We have applied the step-like
model of ion channel switching (\ref{fedi})
in symmetric ($B=5$, $n_0=0.5$) as well 
as asymmetric ($B=5$, $n_0=0.49$) case.
The initial state $\left(P_k(0)\right)$ was chosen to be 
the $50\%$ mixture of the stationary probability distribution
$\left(P_k^{(0)}\right)$ and 
$50\%$ of narrow-peaked distribution,
concentrated in secondary maximum of the stationary distribution
(see. Fig.~\ref{stadis}).
Two characteristics periods have been identified in 
the time evolution of $m=400$ channels 
membrane, in both symmetric and asymmetric case: 
$(i)$ a fast diffusion from the initial conditions
to metastable solution (see Fig.~\ref{stepi}), 
and $(ii)$ a slow relaxation to the stationary solution.
Both of these features are shown in Fig.~\ref{stepi}. 
Our quantitative observations for the membranes with different
numbers of channels $m$ can be briefly reported as follows:  
the time of diffusion ({\em i}) was found to be approximately 
constant for $m=100\div 400$, whereas the life time of the metastable
solution ({\em ii}) grows rapidly with $m$. Moreover,
for large systems, the shape of the probability distribution 
remains exactly constant for a long period of time,
i.e. $\tau=10^2\div 10^4$ for $m=400$, that correspond
to the real time approx. $0.2\div 20\mbox{ s}$ (see Table \ref{kexp}). 

\begin{figure}[b]
\unitlength=0.01\textwidth
\begin{picture}(100,70)
\put(0,19.1){\epsfxsize=\textwidth\epsfbox{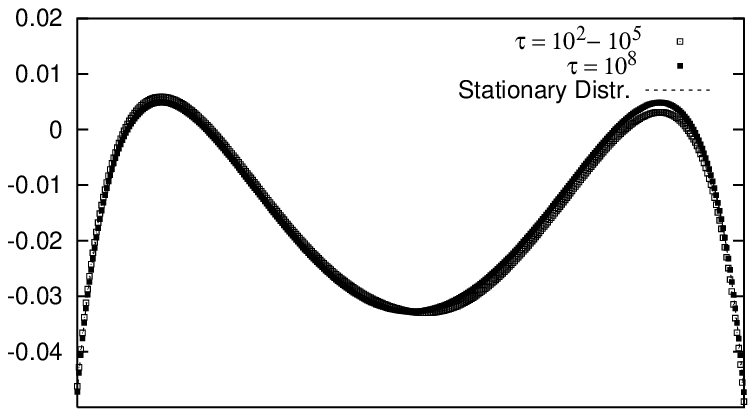}}
\put(0,-12){\epsfxsize=\textwidth\epsfbox{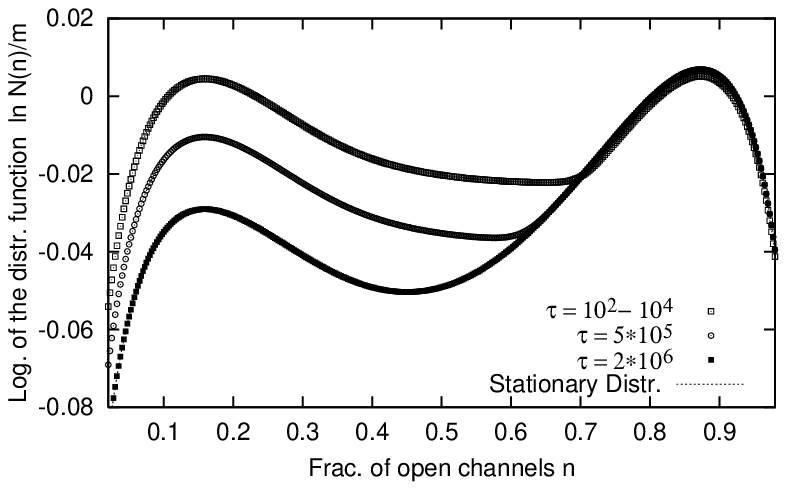}}
\end{picture}
\caption{
The metastable probability distribution 
for $m=400$ channels membrane 
obtained from numerical {\em exact diagonalization} technique, 
(datapoints does not change visibly in given time intervals)
and its relaxation to
the stationary solution for $\tau\rightarrow\infty$ 
(also presented).
The quasi-continuous probability distribution function 
is defined as $P(n)=mP_k$, where $n=k/m$. 
{\em Above}: symmetric step-like model of ion channel
switching ($B=5$, $n_0=0.5$), {\em below}: an asymmetric case
($B=5$, $n_0=0.49$).}
\label{stepii}
\end{figure}

\begin{figure}[b]
\unitlength=0.01\textwidth
\begin{picture}(100,75)
\put(0,23){\epsfxsize=\textwidth\epsfbox{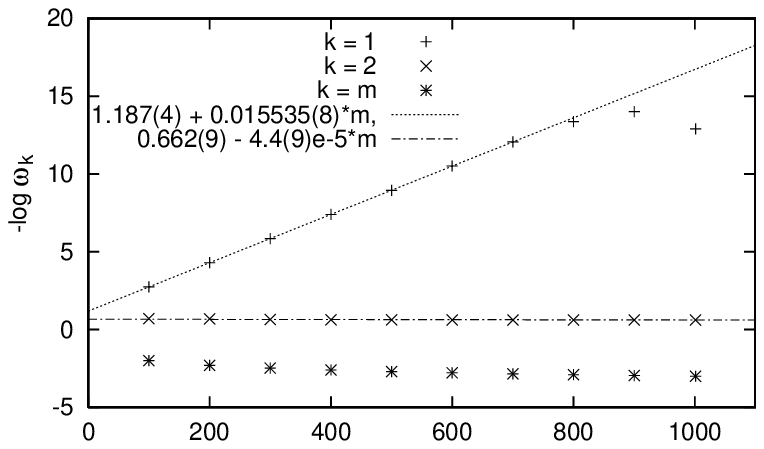}}
\put(0,-12){\epsfxsize=\textwidth\epsfbox{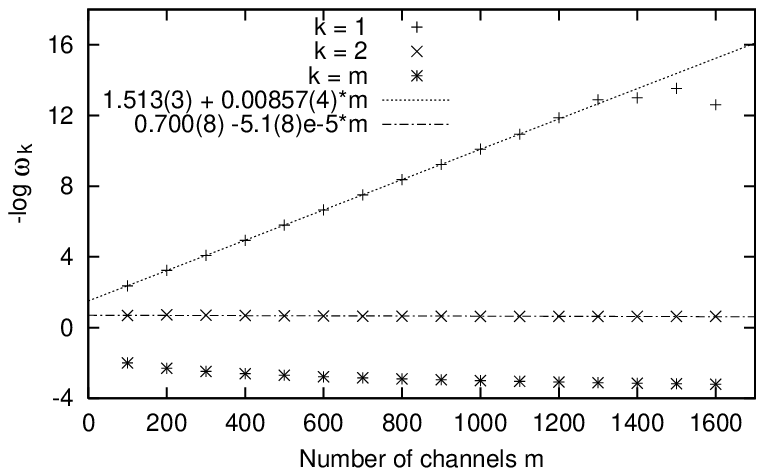}}
\end{picture}
\caption{
Spectrum of eigenvalues obtained from the numerical 
{\em exact diagonalization} approach, 
as a function of the number of channels $m$. 
The first ($\omega_1$), the second ($\omega_2$)
and the last ($\omega_m$) excited eigenvalues are shown, 
all the others are lying between $\omega_2$ and $\omega_m$
(except from $\omega_0=0$). To make interpretation
in terms of decay time easier, $-\log\omega_k$ are 
presented.
{\em Above}: symmetric step-like model of ion channel
switching ($B=5$, $n_0=0.5$), {\em below}: an asymmetric case
($B=5$, $n_0=0.49$).}
\label{specom}
\end{figure}

According to these results, one can expect
a broad window in the spectrum of eigenvalues 
$\left(\omega_k\right)$ of the dynamical system ({\ref{dnkt}}).
Spectra presented in Fig.~\ref{specom} confirm clearly
this hypothesis, for both the symmetric and the asymmetric cases. 
Moreover, first excited-state eigenvalue $\omega_1$ increases 
exponentially with the number of channels $m$, 
(except of the numerical round-off errors, which become
significant for $\omega_1<10^{-12}$),
whereas all the other excited-state eigenvalues lie
in approximately constant (for different $m$) band
bordered by the eigenvalues $\omega_2$ and $\omega_m$. 
This simple picture proves, that for the $m$-s of order
few hundreds and larger, the dynamics of the probability
distribution $\left(P_k\right)$ is {\em completely determined}
by first eigenvalue $\omega_1$ and associated eigenvector
$P^{(1)}=\left(P_k^{(1)}\right)$, for times longer than $\tau\approx 100$
(that correspond to the real time approx. $0.2\mbox{ s}$).

\begin{table}[t]
\caption{
Brief comparison of the decay-time
defined by first excited eigenvalue $1/\omega_1$
with trapping times $T_{\left<n\right>}$ and $T_{Sh}$
(see Fig.~\ref{trapn} and \ref{trapsh} for definitions).
Sample results for {\em symmetric} ($B=5$, $n_0=0.5$)
and {\em asymmetric} ($B=5$, $n_0=0.49$)
step-like models of ion channel switching are provided.}
\centering
\begin{tabular}{|c|c|c|c|c|}
\hline\hline
$n_0$ & $m$ & $-\log\omega_1$ & $\log T_{\left<n\right>}$ 
 & $\log T_{Sh}$ \\ \hline 
0.5 & 100 & 2.748 & 2.540 & 2.392 \\
 & 200 & 4.293 & 4.106 & 3.968 \\
 & 400 & 7.397 & 7.230 & 7.091 \\
\hline
0.49 & 100 & 2.364 & 2.195 & 2.110 \\
 & 200 & 3.236 & 3.057 & 2.948 \\
 & 400 & 4.942 & 4.768 & 4.653 \\
\hline\hline
\end{tabular}
\label{traptab}
\end{table}

\begin{figure}[b]
\unitlength=0.01\textwidth
\begin{picture}(100,70)
\put(0,19.1){\epsfxsize=\textwidth\epsfbox{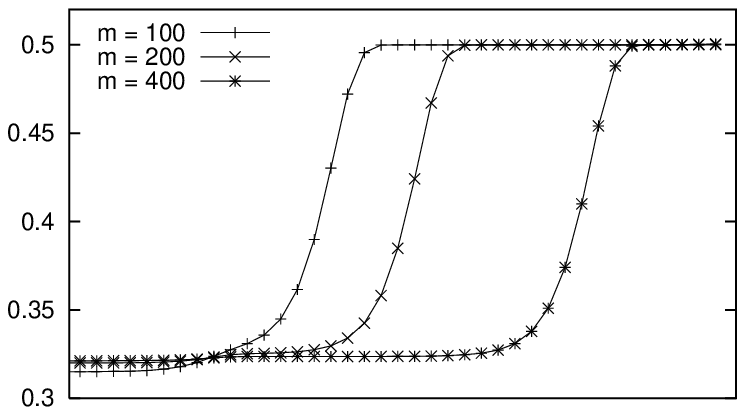}}
\put(0,-12){\epsfxsize=\textwidth\epsfbox{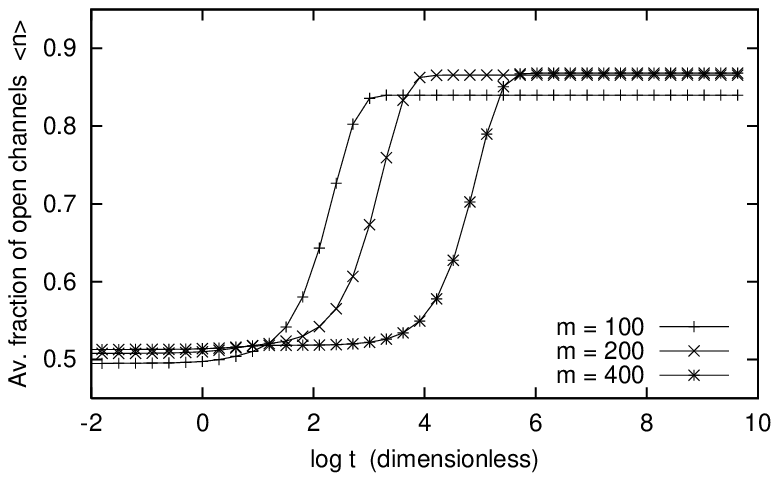}}
\end{picture}
\caption{
Average fraction of open channels 
$\left<n\right>=\int dn nP(n)$, as a function of time
for membranes with $m=100\div 400$ channels with 
symmetric ($B=5$, $n_0=0.5$, {\em above})
and asymmetric ($B=5$, $n_0=0.49$, {\em below}) step-like
model of ion channel switching. 
The quasi-continuous probability distribution function 
is defined as $P(n)=mP_k$, where $n=k/m$. 
The last inflection points on these plots define
{\em trapping times} $T_{\left<n\right>}$, 
collected in Table \ref{traptab}.}
\label{trapn}
\end{figure}

\begin{figure}[b]
\unitlength=0.01\textwidth
\begin{picture}(100,70)
\put(0,19.1){\epsfxsize=\textwidth\epsfbox{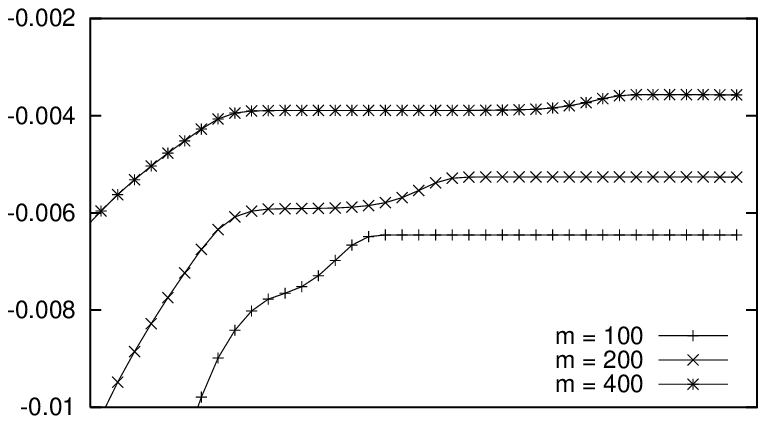}}
\put(0,-12){\epsfxsize=\textwidth\epsfbox{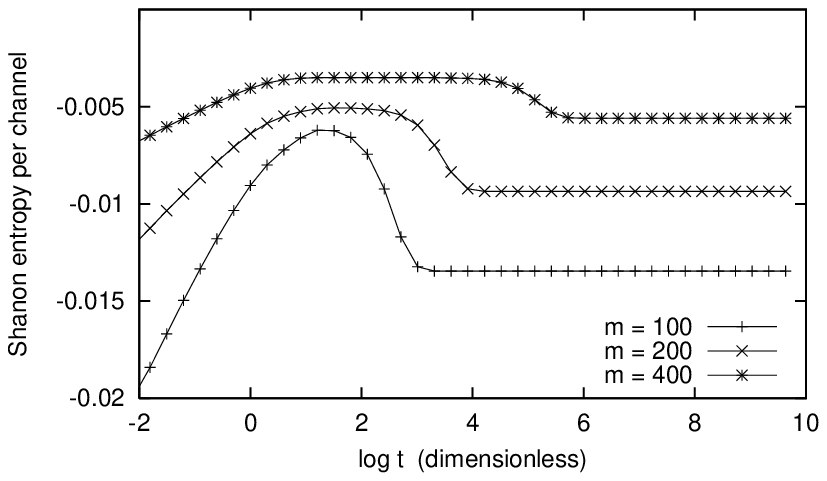}}
\end{picture}
\caption{
Shanon entropy per channel $S_{Sh}/m$, where 
$S_{Sh}=\int dn P(n)\ln P(n)$, as a function of time
for membranes with $m=100\div 400$ channels with 
symmetric ($B=5$, $n_0=0.5$, {\em above})
and asymmetric ($B=5$, $n_0=0.49$, {\em below}) step-like
model of ion channel switching. 
The quasi-continuous probability distribution function 
is defined as $P(n)=mP_k$, where $n=k/m$. 
Last inflection points on these plots define
{\em trapping times} $T_{Sh}$, 
gathered Table \ref{traptab}.}
\label{trapsh}
\end{figure}

A further numerical evidence for this statement 
is provided with Table \ref{traptab}, as well as
Fig.~\ref{trapn}. and \ref{trapsh}. Let us define
the trapping times $T_{\left<n\right>}$ and $T_{Sh}$
as a position of the last inflection points (before 
the system reaches an equilibrium) 
of average fraction of open channels $\left<n\right>$
and the Shannon entropy $S_{Sh}$, respectively 
(see Fig.~\ref{trapn} and \ref{trapsh} for details).
By looking at Table \ref{traptab}, one can find 
the trapping times $T_{\left<n\right>}$ and $T_{Sh}$
to be in the same order as corresponding decay
times defined by first excited eigenvalue $1/\omega_1$, 
i.e. $T_{Sh}\approx 10^{-0.3}/\omega_1\approx 1/2\omega_1$, 
where the factor $1/2$ could be simple explained by representing the
probability distribution in the form:
$P(n)=P_0(n)+\delta(n)$, and calculating the Shanon entropy in 
this manner: 
\begin{equation}
  \label{sshdel}
  S_{Sh} = \int dn P(n)\ln P(n) = 
  S_0 - \frac{1}{2}\int dn\frac{\delta^2}{P_0(n)} 
  + {\cal O}(\delta^3) 
  \approx S_0 + \mbox{const }e^{-2\omega_1\tau},
\end{equation}
where approximation works for $\tau\gg 1/\omega_2$
and $S_0$ is the value of Shanon entropy for the
stationary distribution $P_0(n)$. 
One remarkable feature of the Shanon entropy $S_{Sh}$ for 
the membrane is that its value for the stationary distribution
is not always maximal, for same cases entropy $S_{Sh}$ 
can decrease as a function of time (see e.g. Fig.~\ref{trapsh}). 
This is caused by the fact, that we are studying a dissipative
system, connected to the environment (see Sec.~\ref{model}), 
whereas only the {\em total} entropy of the system together with its 
environment grows in time, not necessarily the {\em fractional}
entropy of the membrane (i.e. Shanon entropy). 

Observations discussed in this section, 
particularly the simple structure of the 
eigenvalue spectrum (presented on Fig.~\ref{specom}),
gave us the motivation for some {\em approximate} approach 
for the first-excited  eigenvalue $\omega_1$, 
that is discussed in the next section.

\section{Perturbative analysis and discussion}
\label{pertu}

Let us consider the membrane with large 
number of channels ($m=200\div 1000$, e.g.)
and a {\em metastable} solution for the fraction
of open channels $n_M$ 
clearly separated from the {\em stable} $n_S$ 
(in other words: secondary and primary
maximum of the stationary probability distribution, 
see i.e. Fig.~\ref{stadis}.) by an {\em unstable}
point $n_U$ (and corresponding minimum
of the stationary probability distribution, 
see Fig.~\ref{stadis}, e.g.).
For this case, according to the broad window in the
spectrum of eigenvalues found 
in Sec.~\ref{dyna}.\ (see Fig.~\ref{specom}.),
the distribution function $P(n)$ relax 
nearly independently in sections
$n<n_S$ and $n>n_S$ after a time
defined by {\em second} excited eigenvalue $\sim 1/\omega_2$,
and then, after a time of order $\sim 1/\omega_1$
(where $\omega_1$ is {\em first} excited eigenvalue)
the probability start to flow through
the unstable point $n_U$, what let
the distribution function $P(n)$ to relax to the
stationary solution $P_0(n)$, 
see Fig.~\ref{stepi}-\ref{stepii}.
Motivated by this brief summary of the results presented in Sec.~\ref{dyna}.,
we discuss now perturbative approach in terms of weak coupling
at the point $n_U$.

Taking into account the probability flows
between point $k_U-1$ and $k_U$, where $k_U\equiv mn_U$
(see Fig~\ref{nflow}), we obtain the equation 
expressing total balance of the probability
in the system in the form
\begin{equation}
  \label{dtdelu}
  \frac{d}{d\tau}\int^{n_U}_0P(n,\tau)dn=
  n_U(1-n_U)\Delta_U(\tau),
\end{equation}
where $\Delta_U$ is defined as a difference
$\Delta_U\equiv P(n_U)-P(n_U-\Delta n)$, $\Delta n=1/m$
and $\tau$ is an dimensionless time (\ref{ltt}).
Whereas the eq.\ (\ref{dtdelu}) has a general
character, for further discussion 
the integral $\int^{n_U}_0P(n)dn$ and difference $\Delta_U$
need to be approximated for $\tau\gg 1/\omega_2$. 
To establish such an approximation,  
one can define the decomposition of
the stationary probability distribution
$P_0=P_L+P_R$, where
\begin{equation}
  \label{nlnr}
  P_L(n)=\left\{\begin{array}{cl}
  P_0(n) & \mbox{for  } n<n_U \\
  0 & \mbox{elsewhere}
  \end{array}\right.,\ \ \ \ \ \ 
  P_R(n)=\left\{\begin{array}{cl}
  P_0(n) & \mbox{for  } n\geq n_U \\
  0 & \mbox{elsewhere}
  \end{array}\right. .
\end{equation}

The representation (\ref{nlnr}) and a brief discussion
of the exact-diagonalization results, presented at
the beginning of this section, brought us to an ansatz:
\begin{equation}
  \label{aeta}
  P_{\eta}(n,\tau)=P_0(n)+\eta(\tau)\left[
  P_L(n)-c_RP_R(n)\right],
\end{equation}
where $\eta(\tau)$ is a function of time, that
we expect to behave as $\eta\sim\exp(-\omega_1\tau)$
for $\tau>>1/\omega_2$, 
and $c_R$ is the coefficient defined to
save the normalization $\int^1_0P(n,1)dn=1$, 
$c_R\equiv P_L/(1-P_L)$, where 
$P_L=\int^{n_U}_0P_0(n)dn$.
After substituting (\ref{aeta}) to the
eq.\ (\ref{dtdelu}), one can write the 
differential equation for the function $\eta(\tau)$: 
\begin{equation}
  \label{dteta}
  \frac{d\eta}{d\tau}=
  -\frac{n_U(1-n_U)}{P_L(1-P_L)}P_U\eta,
\end{equation}
where $P_U$ is the value of the stationary
probability distribution function in point $n_U$,
$P_U\equiv P_0(n_U)$. 
The solution of the eq.\ (\ref{dteta}) is the
simple exponential decay 
$\eta(\tau)=\eta(0)\exp(-\omega_{\eta}\tau)$,
where the decay constant
\begin{equation}
  \label{ometa}
  \omega_{\eta}=\frac{n_U(1-n_U)}{P_L(1-P_L)}P_U
\end{equation}
is an approximation of first excited eigenvalue $\omega_1$.
This approximation is essentially equivalent
to the first-order perturbation correction to the $\omega_1$, 
if we consider the connection in point $n_U$ as a perturbation.

Moreover, one can consider the stationary probability 
distribution in the limit form, for large $m$ (\ref{nexpf}), 
and make the Gaussian approximation to calculate 
all the integrals. The practical formula, obtained in this
manner 
\begin{equation}
\label{logometa}
  \omega_{\eta}=-m(f_M-f_U)\log e + \frac{1}{2}\log m \
  + {\cal O}(1),
\end{equation}
where $f_M$ and $f_U$ are values of the function $f(n)$
(see Sec.\ \ref{stat}) in point $n_M$ and $n_U$, respectively,
can be compared indirectly with numerical 
exact-diagonalization results. 
Although the coefficient 
in the linear part of the eq. (\ref{logometa})
$(f_M-f_U)\log e = 0.01557$ and $0.008597$, 
for symmetric ($B=5$, $n_0=0.5$) and
asymmetric ($B=5$, $n_0=0.49$) Fermi-Dirac 
model of the ion channel switching,
approximately agree with corresponding numerical values
obtained from exact-diagonalization (see Fig.\ \ref{specom}), 
the term $\log m/2$ simply does not exists in numerics.
This clear contradiction seems to be
the immanent disadvantage of the perturbation approach
to our system, that simply produce too large value of
the step $\Delta_U$ in eq.\ (\ref{dtdelu}). 
Moreover, it cannot be repaired in any
perturbation scheme, because, what was shown numerically, the
second perturbation correction to the
first excited eigenvalue $\omega_1$ 
produce nonphysical negative results. 

To avoid the problem mentioned above, one can derive the 
diffusion equation for $n\approx n_U$:
\begin{equation}
\label{ndiff}
  \frac{\partial P}{\partial t} =
  \frac{n_U(1-n_U)}{m}\frac{{\partial}^2P}{\partial n^2}, 
\end{equation}
find its Green function 
\begin{equation}
\label{greendiff}
  G(n,\tau)=\sqrt{\frac{m}{4\pi\tau n_U(1-n_U)}}
  \exp{\left[-\frac{mn^2}{4\tau n_U(1-n_U)}\right]},
\end{equation}
to take the ansatz (\ref{aeta}) as the initial 
condition for the eq.\ (\ref{ndiff}) and {\em frozen} it after
a time ${\tau}_{\rm diff}$: 
\begin{equation}
\label{afroze}
  P_{\eta}^{\rm froze}(n,\tau_{\rm diff})=
  \int dn'G(n-n',\tau_{\rm diff})P_{\eta}(n',\tau).
\end{equation}
Then, the value of the step $\Delta_U$ in eq.\ (\ref{dtdelu}) 
is multiplied by factor $2G(0,\tau_{\rm diff})\Delta n$, where 
$G(0,\tau_{\rm diff})$ is the value of Green's function (\ref{greendiff})
at point $n=0$ and $\Delta_n=1/m$. That brought us to the new 
approximation of first excited eigenvalue
\begin{equation}
\label{omfroze}
  \omega_{\eta}^{\rm froze}=\frac{2n_U(1-n_U)}{P_L(1-P_L)}
  G(0,\tau_{\rm diff})P_U\Delta n=
  \sqrt{\frac{n_U(1-n_U)}{\pi m\tau_{\rm diff}}}
  \frac{P_U}{P_L(1-P_L)}.
\end{equation}
The factor $\sqrt{1/m}$ appeared in eq.\ (\ref{omfroze})
cancel the logarithmic term in (\ref{logometa}). 

In so-called {\em frozen diffusion} approach, presented above, 
the diffusion time $\tau_{\rm diff}$ 
is an free parameter, adjusted
to agree with constants in formula for $\log\omega_1$ 
obtained from exact diagonalization. 
Although the {\em frozen diffusion} approach
produce the results agreeable with the numerics, 
the physical reasons for introducing
parameter $\tau_{\rm diff}$ are not clear at this stage; 
the problem needs a~further study, possibly with 
utilising the {\em Real-Space Renormalization Group Technique}.

\section*{Summary}

We derived the {\em detail-balance} equations 
for stationary probability distribution 
of the number of open ion channels in the 
cell membrane. This equation was 
used to compare the extrema of
the stationary probability distribution for large $m$-channel system
with corresponding stationary solutions 
of the deterministic dynamical equation for $m\rightarrow\infty$,
and to discuss the fluctuation around them for finite $m$. 

The so-called {\em step-like} model of ion 
channel  switching, based on that studied by Hodgkin and Huxley \cite{hohu}, 
was briefly discussed in relation to
the experimental data, applied to obtain an analytical
solution of the detail-balance equation, 
qualitative discussion of the two classes 
of phase transitions in the system, and
numerical {\em exact-diagonalization} calculations
of the decay time of the metastable state of the system. 
The exact diagonalization results were shown to disagree 
with the first-order perturbations. 
Moreover, the perturbation scheme produces unphysical results 
in the second order. 
This contradiction shows essentially a non-perturbative character
of coupled ion-channel system.
The proposed {\em frozen diffusion} approach,  
with an additional adjustable parameter, was
shown to explain the exact diagonalization 
results.

\section*{Acknowledgement}

Author would like to thank Prof.\ S.~B.~Fahy for
supervising the work at the early stage.
I am also grateful to Prof.\ K.~Ro\'sciszewski and, 
particularly, to Prof.\ J.~Spa{\l}ek
for discussions and critical reading of the manuscript. 
The project was mainly completed during the 3-months stay 
at the University College Cork, Ireland, that was supported 
by TEMPUS Mobility Grant No. S-JEP-12249-97.  
The support of the KBN Grant No.\ 2P03B~092~18 is also appreciated.

\appendix
\section{Fluctuations around the stationary points
	obtained from the detailed-balance approach}
\label{detbal}

In this appendix, we present the approximate derivation of the small 
fluctuations of the number of open channels around the stationary points 
for general model of ion channel switching. Let us apply the detailed balance 
equation (\ref{detbalnn}) to calculate the ratio of the probability 
distribution $P_{k-\Delta}/P_k$ for points in distance $\Delta$:
\begin{equation}
  \label{nkdelnk}
  \frac{P_{k-\Delta}}{P_k}=
  \prod^{\Delta-1}_{j=0}\frac{P_{k-j-1}}{P_{k-j}}=
  \prod^{\Delta-1}_{j=0}\frac{k-j}{m-k+j+1}
  \frac{1-p_{k-j}}{p_{k-j-1}},
\end{equation}
where probability of open channel 
$p_k\equiv p_{open}(k/m)$. 
For $\Delta/m<<1$ ({\em small fluctuations}),
which supposition would be verified {\em a posteriori}, 
one can approximate the probabilities in 
eq.\ (\ref{nkdelnk})
$$
  p_{k-j}\approx p_k-jp_k' = p_k(1-\alpha j),
$$
where $p_k'=p_{open}'(j/m)/m$, and 
\begin{equation}
\label{alphapp}
  \alpha=\frac{p_k'}{p_k}=
  \frac{1}{m}\frac{p_{open}'(k/m)}{p_{open}(k/m)}<<1;
\end{equation}
and analogically
$$
  1-p_{k-j}\approx 1-p_k(1-\alpha_j)=
  (1-p_k)(1+\beta j),
$$
where:
\begin{equation}
\label{betapp}
  \beta = \alpha\frac{p_k}{1-p_k}=
  \frac{p_k'}{1-p_k}<<1.
\end{equation}
Subsequently, eq.\ (\ref{nkdelnk}) can be 
rewrite in the approximate form: 
\begin{equation}
\label{binobi}
  \frac{P_{k-\Delta}}{P_k}\approx
  \frac{k!}{(k-\Delta)!}
  \frac{(m-k)!}{(m-k+\Delta)!}
  \left(\frac{1-p_k)}{p_k}\right)^{\Delta}
  \times\Pi_{\Delta},
\end{equation}
where $\Pi_{\Delta}$ is the non-binomial factor 
in ratio $P_{k-\Delta}/P_k$: 
\begin{equation}
\label{pidel}
  \Pi_{\Delta}=\prod^{\Delta-1}_{j=0}
  \frac{1+\beta j}{1-\alpha (j+1)}\approx
  \prod^{\Delta-1}_{j=0}
  \left[1+(\alpha+\beta)j\right]\approx
  e^{(\alpha+\beta)\Delta^2/2},
\end{equation}
where we skipped the terms
of order $\Delta/m$.

Furthermore, for stationary point $p_k=k/m\equiv n_*$
(see Sec.\ \ref{dete}). 
So that, using Stirling approximation 
$\ln k!\approx k\ln k-k$, one can obtain 
from eq.\ (\ref{binobi}):
$$
  \ln\frac{P_{k-\Delta}}{P_k}\approx
  -\frac{\Delta^2}{2mn_*(1-n_*)}+
  \ln\Pi_{\Delta}.
$$
After substituting eq.\ (\ref{pidel}) and,
subsequently (\ref{alphapp}) and (\ref{betapp}),
we finally obtain:
\begin{equation}
\label{nkdelnkstat}
  \ln\frac{P_{k-\Delta}}{P_k}\approx
  -\frac{1-p_*'}{2mn_*(1-n_*)}\Delta^2,
\end{equation}
where $p_*'\equiv p_{open}'(n_*)$.
Eq.\ (\ref{nkdelnkstat}) let us read the mean-square
fluctuation of the fraction of open channels 
$\delta n\equiv\Delta/m$ around the stationary point $n_*$:
\begin{equation}
\label{deltn}
  \left<(\delta n)^2\right> =  
  \frac{1}{m}\frac{n_*(1-n_*)}{1-p_*'},
\end{equation}
that confirms our starting 
supposition $\Delta/m<<1$ for large $m$. 
Moreover, eq.\ (\ref{deltn}) let
us to identify the value $p_*'=1$, 
previously distinguishing between types of
stationary solutions (see Sec.\ \ref{dete}.), 
with {\em critical point} 
in terms of 2-nd order phase transitions, 
for which $\delta n\rightarrow\infty$. 
Additionally, for quasi-continuous probability distribution
function $P(n)$, defined by the relation $P_k=P(n)\Delta n$
(cf.\ Sec.\ \ref{stat}.) one can write (for $n\approx n_*$):
\begin{equation}
\label{ngauss}
  P(n)=P(n_*)\exp\left[
  -m\frac{1-p_*'}{n_*(1-n_*)}(n-n_*)^2,
  \right]
\end{equation}
what clearly identifies {\em stable} and 
{\em unstable} solutions of the relation $n_*=p_{open}(n_*)$
with, respectively maxima and minima of the probability distribution. 
Eq.\ (\ref{ngauss}) motivates the ansatz $P\sim e^{mf(n)}$
in Sec.\ \ref{stat}.


\begin{thebibliography}{99}
\bibitem{further}
The further description requires 
phenomenological models for parameters $B=B(V,R/R_m,\Theta)$, 
$n_0=n_0(V,R/R_m,\Theta)$, and will not be presented here.
\end{thebibliography}
\end{document}